\documentclass[reprint, amssymb, amsmath,aip,apl]{revtex4-1}

\usepackage{graphicx}
\usepackage{amssymb}
\usepackage{epstopdf}
\usepackage{upgreek}
\usepackage{dcolumn}
\usepackage{bm}
\expandafter\ifx\csname package@font\endcsname\relax\else
 \expandafter\expandafter
 \expandafter\usepackage
 \expandafter\expandafter
 \expandafter{\csname package@font\endcsname}%
\fi

\newcommand{\be}{\begin{eqnarray}}
\newcommand{\ee}{\end{eqnarray}}
\newcommand{\bfig}{\begin{figure}}
\newcommand{\efig}{\end{figure}}

\begin{document}

\title{Single-sideband modulator for frequency domain multiplexing of superconducting qubit readout}
\author{Benjamin J. Chapman}
\email{benjamin.chapman@colorado.edu}
\affiliation{JILA, National Institute of Standards and Technology and the University of Colorado, Boulder, Colorado 80309, USA}
\affiliation{Department of Physics, University of Colorado, Boulder, Colorado 80309, USA}
\author{Eric I. Rosenthal}
\affiliation{JILA, National Institute of Standards and Technology and the University of Colorado, Boulder, Colorado 80309, USA}
\affiliation{Department of Physics, University of Colorado, Boulder, Colorado 80309, USA}
\author{Joseph Kerckhoff}
\altaffiliation{Current address: HRL Laboratories, LLC, Malibu, CA 90265, USA}
\affiliation{JILA, National Institute of Standards and Technology and the University of Colorado, Boulder, Colorado 80309, USA}
\affiliation{Department of Physics, University of Colorado, Boulder, Colorado 80309, USA}
\author{Leila R. Vale}
\affiliation{National Institute of Standards and Technology, Boulder, Colorado 80305, USA}
\author{Gene C. Hilton}
\affiliation{National Institute of Standards and Technology, Boulder, Colorado 80305, USA}
\author{K.~W. Lehnert}
\affiliation{JILA, National Institute of Standards and Technology and the University of Colorado, Boulder, Colorado 80309, USA}
\affiliation{Department of Physics, University of Colorado, Boulder, Colorado 80309, USA}
\date{\today}

\begin{abstract}
We introduce and experimentally characterize a superconducting single-sideband modulator compatible with cryogenic microwave circuits, and propose its use for frequency domain multiplexing of superconducting qubit readout. The monolithic single-quadrature modulators that comprise the device are formed with purely reactive elements (capacitors and Josephson junction inductors) and require no microwave-frequency control tones.  Microwave signals in the 4 to 8 GHz band, with power up to -85 dBm, are converted up or down in frequency by as much as 120 MHz. Spurious harmonics in the device can be suppressed by up to 25 dB for select probe and modulation frequencies.
%We introduce and experimentally characterize a superconducting single-sideband modulator (SSBM) compatible with cryogenic microwave circuits, and propose its use for frequency domain multiplexing of superconducting qubit readout. The monolithic single-quadrature mixers that comprise the device are formed with purely reactive elements and require no microwave-frequency control tones. Operation is demonstrated across the 4 to 8 GHz band for microwave tones with power up to -85 dBm. Signals converted by up to 20 MHz exceed other spectral components by 20 dB or more at some carrier frequencies.
\end{abstract}

\maketitle
Recent advances have allowed many groups to demonstrate superb control over one or several superconducting qubits. This is seen, for example, in the creation and transmission of Fock and cat state superpositions\cite{wang:2016,pfaff:2016}, %the generation of remote entanglement between pairs of qubits\cite{narla:2016}, 
or the ability to continuously tune both the strength\cite{hatridge:2013} and axis\cite{hacohen:2016,vool:2016} of a quantum non-demolition\cite{wallraff:2004} measurement. %near-arbitrary Fock and cat states
Efforts to combine multiple superconducting qubits into a (more coherent) logical qubit are progressing in parallel.\cite{kelly:2015,ofek:2016}  %As such, quantum information processing with superconducting qubits is approaching a new stage\cite{devoret:2013}.  
As  algorithms on multiple logical qubits become experimentally realizable, the challenge of scaling readout for many-qubit systems grows in relevance. 

%Efforts to utilize multiple superconducting qubits for error correction are progressing in parallel.  The construction of a logical qubit more coherent than its constituent physical qubits is a major achievement in this vein.\cite{kelly:2015,ofek:2016}  As such, quantum information processing with superconducting qubits is approaching a new stage\cite{devoret:2013}.  As  algorithms on multiple logical qubits become experimentally realizable, the challenge of scaling readout for many-qubit systems grows in relevance.  

Multiplexing, a signal processing technique in which a transmission medium is shared among multiple signals, is a natural way to combat such a scaling challenge. In this letter, we introduce a single-sideband modulator (SSBM) for cryogenic microwave applications and propose its use for frequency domain multiplexing of superconducting qubit readout. We describe its design, theory of operation, and experimental performance.

%Readout of a superconducting qubit is frequently done with a transmission measurement: when a qubit dresses a microwave cavity, the phase~\cite{blais:2004} or amplitude~\cite{reed:2010} of a transmitted tone can indicate the state of the qubit.  
In a circuit quantum electrodynamics architecture, qubits are readout by measuring the transmission of a cavity that has a resonance frequency that is modified (dressed) by a qubit.\cite{blais:2004} Specifically, the cavity resonance is dressed by an amount that depends on the qubit state.  To-date, multi-qubit experiments have scaled-up this readout protocol directly~\cite{takita:2016} or with frequency domain multiplexing (FDM) hardwired into the measurement architecture.\cite{chen:2012,jerger:2012,kelly:2015}  In direct ``brute-force'' scaling, each qubit channel has a dedicated input and output line.  Conversely, qubits in an FDM architecture can share a single input and output line, and their readout cavities are designed in advance to ensure their dressed resonances are separated by multiple cavity linewidths.%In FDM, the cavities are designed in advance to ensure their dressed resonances are separated by multiple cavity linewidths and are spectrally resolvable.

As the number of readout channels in a multi-qubit measurement rises, direct scaling becomes untenable.  %The cost of the microwave receivers and the finite hardware volume afforded by a dilution refrigerator each speak to the need for scalable multiplexed architectures with cryogenic analog signal processing.  
Recognition of this fact has spurred new proposals for scalable architectures~\cite{brecht:2015,chapman:2016}, as well as a slew of devices for in-fridge signal processing, from non-reciprocal elements~\cite{ranzani:2015,metelmann:2015,kerckhoff:2015,sliwa:2015,lecocq:2016,abdo:2017} to mixers~\cite{bergeal:2010,naaman:2016} and switches.\cite{naaman:2016switch,chapman:2016,pechal:2016}

The ``hard-wired'' FDM approach achieves scalability at the cost of flexibility.  The spectral allotment for each channel is determined prior to fabrication and, unless the readout resonators are designed to be tunable\cite{whittaker:2016}, %typically\cite{tunable}
remains fixed through the measurement process.  This constraint inhibits efficient use of the available bandwidth, as densely packing the spectra allotted for each channel makes nearby channels sensitive to deviations in the design parameters.  It also precludes the use of identical qubit-cavities, which may be desirable in quantum simulation applications.  Finally, the ``hard-wired'' architecture limits the degree to which a many-qubit experiment can be reconfigured without fabrication of new devices.  %It also precludes the dynamica allocation of bandwidth.  This is useful, for example, if different signal-to-noise ratios (averaging times) are desired for different measurement channels.  

There is thus a need for \textit{flexible} multiplexed architectures that \textit{efficiently} utilize the available measurement bandwidth. To this end, we introduce a microwave component for cryogenic analog signal processing, which could enable flexible FDM in a many-qubit measurement. This component is a single-sideband modulator (SSBM), a two-port device that converts incident signals up or down in frequency.  Unlike commercial SSBMs which utilize resistive diodes, in our device the nonlinear elements are purely reactive.  For such systems, the Manley-Rowe relations permit frequency-conversion without dissipation.\cite{manley:1956}  
Moreover, and in contrast to other cryogenic microwave frequency converters\cite{roch:2012,abdo:2013,lecocq:2016}, the device a) utilizes no resonant physics, endowing it with GHz of bandwidth, and b) drives its non-linear elements with RF signal frequencies no greater than several hundred MHz, obviating the need for high-bandwidth GHz control lines.  Finally, the device can be realized as a monolithic integrated circuit in a NbAlO$_x$Nb tri-layer process~\cite{sauvageau:1995,mates:2008}, allowing for high-yield wafer-scale production.  

\begin{figure}[htb]
\begin{center}
\includegraphics[width=1.0\linewidth]{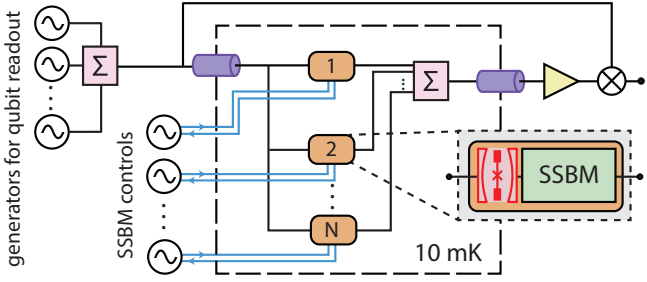}
\caption {Conceptual schematic for the proposed frequency-domain-multiplexed readout of N superconducting qubit/cavity systems (red box) using N single-sideband modulators (SSBM, green rectangle).  The SSBMs frequency-convert tones exiting the qubit read-out cavities for efficient and dynamically reconfigurable frequency domain multiplexing.  As the frequency of the readout tone of each qubit/cavity is converted to a unique portion of the measurement bandwidth, all channels can be readout simultaneously with a single microwave receiver.}
\label{fig:fig0}
\end{center}
\end{figure}

A proposed use for the device is shown in Fig.~\ref{fig:fig0}. By positioning a SSBM at the output of the readout resonator's strongly coupled port of the qubit/cavity system in each measurement channel, a single microwave line can be used to readout all the channels simultaneously in an FDM scheme.  Measurement bandwidth can be dynamically allocated among channels by converting each transmitted tone to its assigned band.  This ability to assign spectrum in-situ facilitates efficient use of the available measurement bandwidth, as there is no risk of spectral overlap from uncertainty in fabrication parameters.  Furthermore, it allows for the use of FDM even if the dressed cavity frequencies are spectrally irresolvable.  Although other multiplexing proposals for time and code-domain schemes\cite{chapman:2016} share these benefits, the flexible FDM architecture depicted in Fig.~\ref{fig:fig0} distinguishes itself by requiring no timing coordination between channels.

%n this letter, we introduce an SSBM intended for FDM of superconducting qubit readout.  We describe its design, theory of operation, and evaluate its performance experimentally. 

\begin{figure}[htb]
\begin{center}
\includegraphics[width=1.0\linewidth]{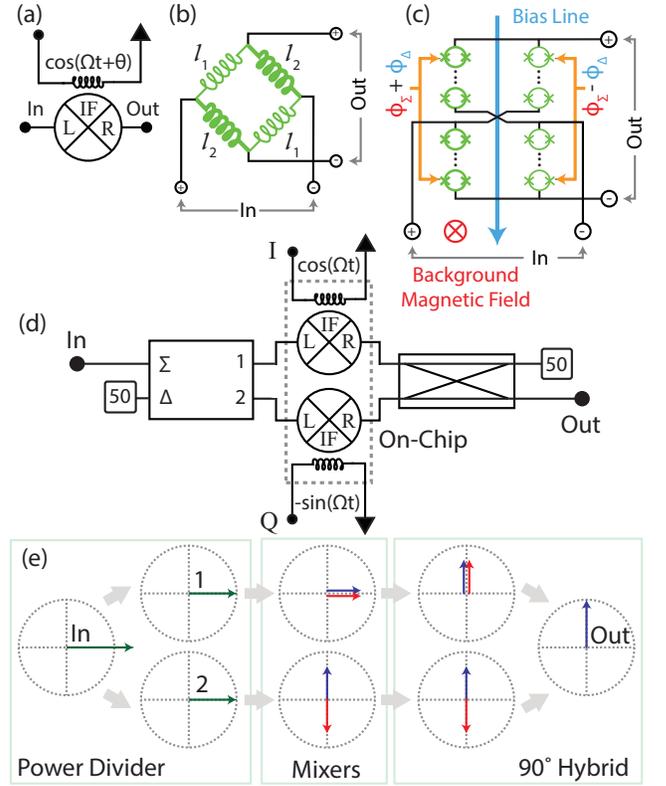}
\caption {(a) Double-balanced modulator used to construct the SSBM, realized with a Tunable Inductor Bridge (b) built from two pairs of nominally identical inductors $l_1$ and $l_2$ arranged opposite one-another.  Input and output ports are defined by the left-right and top-bottom nodes of the bridge. Transmission through the bridge scales with the inductor imbalance $l_1 - l_2$.  (c) A Tunable Inductor Bridge realized with arrays of SQUIDs. Bridge imbalance is modulated with a static background magnetic flux $\phi_\Sigma$ applied with an off-chip magnetic coil and a time-dependent gradiometric flux $\phi_\Delta$ applied with an on-chip bias line.  (d) Schematic of an SSBM composed of a power divider, two double-balanced modulators with IF ports driven in quadrature, and a 90-degree hybrid coupler. (e) Phasor representation of signals traveling through an idealized SSBM (color indicates relative frequency) during frequency conversion. %Signal power is first split into the two arms of the network. The signals are then modulated by the TIBs, with on-chip bias lines driven to modulate transmission at the same angular frequency $\Omega$ but with a phase of $\pi/2$ radians in the second double-balanced modulator. 
%Modulated signals are recombined using a hybrid coupler and the output signal is converted up in frequency by $\Omega$.
}
\label{fig:fig1}
\end{center}
\end{figure}

The SSBM is created from a Hartley-type\cite{razavi:1998} in-phase--quadrature (IQ) modulator with I and Q ports driven 90 degrees out of phase.  This IQ modulator is itself built from a pair of double-balanced modulators (Fig.~\ref{fig:fig1}a) realized with inductive bridge circuits (Fig.~\ref{fig:fig1}b),  which we call Tunable Inductor Bridges (TIBs).\cite{chapman:2016}  Two pairs of tunable inductors form the bridge, with inductances $l_1$ and $l_2$.  %Input and output ports are defined by comparing the voltage differences at the east-west and north-south nodes of the bridge. 
Signals couple differentially to the input port (left and right bridge nodes) and output port (top and bottom bridge nodes).  Transmission between these ports is controlled by the imbalance in the bridge $l_1-l_2$.

%Manipulation of that imbalance is accomplished through the use of four flux-tunable inductors arranged in a figure-eight geometry, which tune in concert to balance (or imbalance) the bridge.  
Manipulation of that imbalance is accomplished by arranging the two pairs of flux-tunable inductors in a figure-eight geometry. %, which tune in concert to balance (or imbalance) the bridge. 
Their coordinated tuning is controlled with an off-chip magnetic coil and an on-chip bias line that bisects the figure eight (Fig.~\ref{fig:fig1}c).\cite{kerckhoff:2015} Series arrays of Superconducting Quantum Interference Devices (SQUIDs) are used to realize the flux-tunable inductors.  Arrays are employed in place of individual SQUIDs to dilute the Josephson nonlinearity.\cite{kerckhoff:2015}  %More details about the layout, fabrication, and performance of a TIB are reported elsewhere.\cite{chapman:2016}

To operate the TIB as a double-balanced modulator, we use the flux-control bias line to sinusoidally modulate the transmission through the bridge at angular frequency $\Omega$. %This mixes an input tone at angular frequency $\omega$ into red and blue sidebands at angular frequencies $\omega \pm \Omega$, as shown in Fig.~\ref{fig:fig1}c.  
Borrowing language from the field of nonlinear-microwave elements, we denote the two microwave ports of the TIB as the local oscillator (L) and radio frequency (R) ports of the modulator, and the flux-control line as the intermediate frequency (IF) port.  %This terminology suggests the TIBs be viewed as mixer.  
However, as a TIB has no galvanic connections between its IF and L port (or its IF and R port), we refer to it as a modulator rather than a mixer.  Note also that in contrast to a common way of operating mixers by driving their nonlinear elements with a strong signal in their L ports, the IF port of the TIB actuates the modulation process and signals at the L and R ports can be arbitrarily small.  %Furthermore, it should be noted that the 1 dB compression point of the L and R ports of a TIB mixer is -88 dBm\cite{chapman:2016}, minuscule compared to a commercial mixer.  Fortunately, that power level still exceeds the power in a typical dispersive readout tone by several orders of magnitude\cite{riste:2013}.

We create an IQ modulator by dividing the power of an input tone into two double-balanced modulators and then summing their outputs with a 90-degree hybrid coupler (Fig.~\ref{fig:fig1}d).  In the device reported in this letter, both double-balanced modulators are integrated on a single chip and connected to a commercial power splitter and 90-degree hybrid.  Future versions of the SSBM, however, could also integrate the passive components on-chip.\cite{ku:2011,pechal:2016} %A purely monolithic design would reduce device footprint, eliminate insertion loss from the power divider and hybrid  %(as they could be constructed from superconductors), 
%and suppress path length differences in the two arms of the device far below operation wavelengths.  %Transitioning to a lumped-element regime could also match the instantaneous impedance of the device to 50 Ohms, reducing reflections by 3 dB.% within the device, as will be discussed shortly.

Fig.~\ref{fig:fig1}e provides a phasor representation of a microwave tone as it propagates through the SSBM.  This tone is first split into two arms with equal phase shift, each connected to the L port of a double-balanced modulator.  %In the top arm, a double-balanced modulator with I port driven at angular frequency $\Omega$ modulates the input tone into two tones spectrally shifted from the original by $\pm \Omega$ (blue/red sidebands, respectively).  Simultaneously, in the bottom arm of the network, the second double-balanced modulator's I port is also driven at angular frequency $\Omega$, but with a phase of $\pi/2$ radians.  
Control currents are applied to the IF ports of both modulators, modulating the signal in each arm into two tones spectrally shifted from the original by $\pm \Omega$ (blue/red sidebands, respectively).  The phase of the control current in the lower arm, $\theta$, is advanced $\pi/2$ radians with respect to that in the upper.  
%Finally, at the 90 degree hybrid, the signal in the top arm is rotated by 90 degrees and summed with the unrotated signal in the bottom arm, causing the blue (red) sidebands to constructively (destructively) interfere at the device's output.  The input tone is thus converted up in frequency by $\Omega$.  Generally, the amplitude of the blue sideband scales as $\cos\left((\pi/2 - \theta)/2\right)$, while the red sideband's amplitude scales as $\cos\left((\pi/2 + \theta)/2\right)$.  Here $\theta$ is the phase of the I drive in the lower double-balanced modulator.  The red sideband can therefore be selected by driving the I port of the double-balanced modulator in the bottom arm with a phase of $-\pi/2$ radians.  We denote the I port in the top arm as the I port of the IQ modulator, and the I port of the double-balanced modulator in the bottom arm as the Q port.
Recombination at the 90-degree hybrid causes the blue (red) sidebands in each arm to constructively (destructively) interfere at the device's output.  The input tone is thus converted up in frequency by $\Omega$.  Generally, the amplitude of the red and blue sidebands scales as $\cos\left((\pi/2 \pm \theta)/2\right)$, allowing for selection of the red sideband when $\theta = -\pi/2$ radians.  We denote the IF port in the top arm as the I port of the IQ modulator, and the IF port of the double-balanced modulator in the bottom arm as the Q port.
\begin{figure}[htb]
\begin{center}
\includegraphics[width=1.0\linewidth]{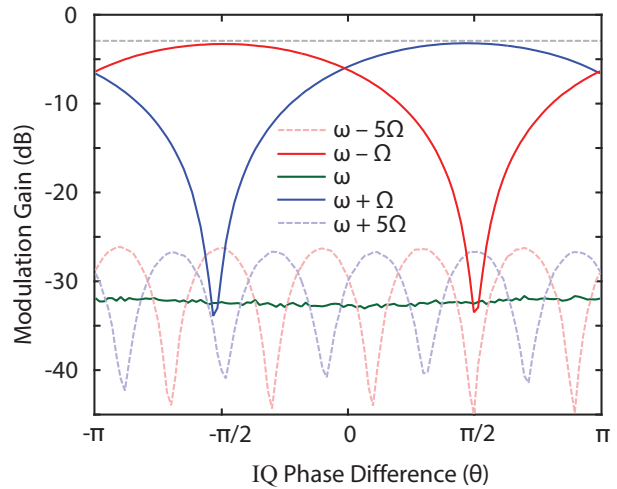}
\caption {Modulation gain of circuit shown in Fig.~\ref{fig:fig1}d, as a function of IQ phase difference $\theta$, with L port driven at $\omega = 2 \pi \times 4$ GHz and transmission modulated at $\Omega = 2 \pi \times$ 3 MHz. Note the 3 dB of power dissipated in the image sideband is not reflected in this measure.  The first upper and lower sidebands are shown in solid blue and red, respectively. They exceed the input frequency (green) by up to 30 dB. Power is also detected at higher harmonics of the modulation frequency.  For the plotted L and I frequencies, the largest harmonic is the fifth (dashed blue and red, respectively). %Transmitted power is normalized to an unmodulated input tone at $\omega$ (see text). 
A gray dashed line at -3 dB is a guide to the eye.  At the operating points $\theta = \pm \pi/2$, the power in the first sideband exceeds all other harmonics by more than 20 dB. 
}
\label{fig:fig2}
\end{center}
\end{figure}

This angular dependence of sideband power is depicted in Fig.~\ref{fig:fig2}, which shows the transmitted power at various frequencies as a function of the IQ phase difference $\theta$.  For this measurement, the SSBM is driven by a microwave tone at $\omega = 2 \pi \times 4$ GHz through its L port while we modulate the transmission across the bridges at $\Omega = 2 \pi \times 3$ MHz.  % with control currents applied to the I and Q ports.  
As this modulation is not purely comprised of a single spectral component at angular frequency $\Omega$, higher harmonics of $\Omega$ at frequencies $\omega \pm n \Omega$, with $n$ an integer greater than 1, are also observed.  We monitored the nearest 16 higher harmonics ($2\leq n\leq9$), and found the $n=5$ harmonic to be the largest at this particular choice of $\omega$ and $\Omega$.  The plot shows the output of the SSBM at the input frequency (green), along with the first upper (blue) and lower (red) sidebands, and the largest of the higher harmonics ($n=5$, dashed red and blue).  More harmonics and phase sweeps at different IQ modulation frequencies are shown in Fig. S3 of the supplementary material.  As $\theta$ is varied between $-\pi$ and $\pi$ we observe that all measured harmonics oscillate with $\theta$ at a rate commensurate with their order.

Two clear operation points are visible at $\theta = \pm \pi/2$ radians, where the first blue (red) sideband is suppressed in favor of the first red (blue) sideband.  The contrast between them exceeds 30 dB at these points, and the contrast between the desired sideband and the nearest harmonic (the 5th, in this case), which we call the sideband contrast, is approximately 23 dB.  

We set the scale of the y-axis in units of dB relative to an unmodulated input tone, which we call modulation gain.  This is done by statically biasing both TIBs to their state of maximal imbalance. We normalize the sideband power in this way to reveal inefficiencies in the modulation process and separate them from the intrinsic insertion loss of the SSBM's constituent components.  %Hardware return and insertion losses will be discussed shortly.  
In  Fig.~\ref{fig:fig2}, the modulation gain in the desired sideband never exceeds -3 dB (dashed gray line) because the TIB in each arm of the interferometer is biased to modulate transmission sinusoidally.  As the device is realized as a distributed network, that means each arm is reflecting half its incident power on average---power which is ultimately dissipated in the power splitter, or directed out the device's input port (return loss).   

%The -3 dB maximum power of the first red and blue sidebands results from time averaging the power in the sinusoidally modulated transmission and then normalizing by the unmodulated tone.  This 3 dB of return loss can be recovered by moving to the lumped element regime with an on-chip implementation, as discussed above.

\begin{figure}[htb]
\begin{center}
\includegraphics[width=1.0\linewidth]{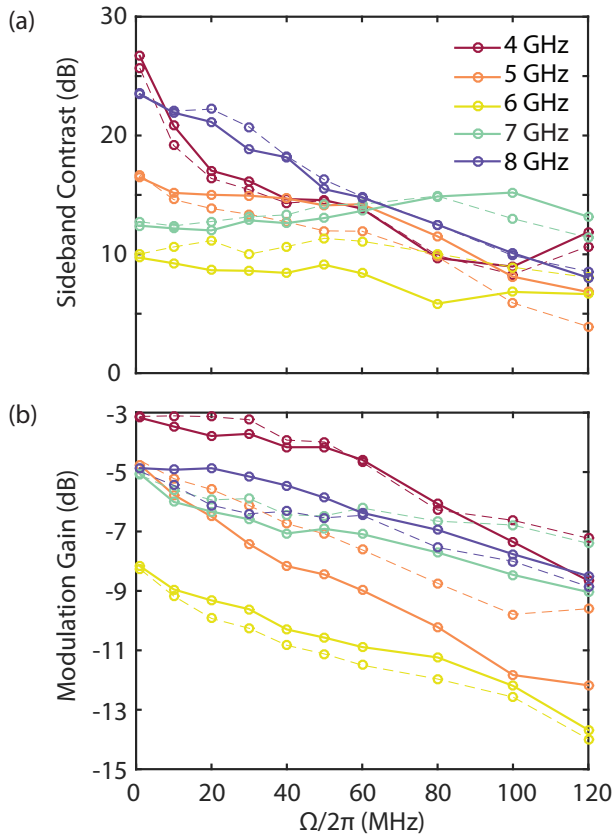}
\caption {%Power difference between the desired sideband and the next largest harmonic (which we denote as sideband contrast) as a function of the I angular frequency $\Omega$. The different traces depict different L frequencies $\omega/(2 \pi)$.  Frequency dependent transmission through the lines has been removed using a bypass switch. (b) Image rejection (contrast between the desired first sideband and the undesired first sideband) over the same range of L and I frequencies. (c) Conversion-gain for up (solid lines) and down (dashed lines) converted tones, as a function of $\Omega$.  Different traces show different L frequencies $\omega/(2\pi)$, all normalized to an unmodulated carrier.  The decrease in conversion-gain with $\Omega$ is a consequence of the control hardware's bandwidth limitations (see supplementary information for more details).
Sideband contrast (a) and modulation gain (b) of the SSBM, as a function of the modulation frequency $\Omega$. The different traces depict different L frequencies $\omega/(2 \pi)$ (color), for selection of the blue (solid lines) and red (dashed lines) sidebands.  %In (b), th are normalized to an unmodulated input tone (see text).
}
\label{fig:fig3}
\end{center}
\end{figure}

%To characterize the device over a broader range of operation frequencies, sideband contrast for a variety of L and I frequencies is plotted in Fig.~\ref{fig:fig3}a.  In the proposed application of the SSBM for FDM of qubit readout, this is an important measure of device performance.  When a single SSBM converts the tone exiting the qubit readout cavity in the $n^\textrm{th}$ channel to its allotted band, other measurement channels may occupy the surrounding spectrum.  Creation of spurious sidebands is then not only a loss mechanism in the $n^\textrm{th}$ channel, but also a source of cross-talk that creates noise in other channels.  
To characterize the device over a broader range of operation frequencies, sideband contrast for a variety of L and I frequencies is plotted in Fig.~\ref{fig:fig3}a.  In the proposed application of the SSBM for FDM of qubit readout, this is an important measure of device performance: when the  measurement spectrum is densely filled, spurious sidebands are a source of cross-talk among channels.  

%To calibrate the conversion gain of our device, we configure the measurement with a bypass switch that allows signals to be routed around the SSBM.  In the current implementation, this gain never exceeds -6 dB, because half the power is dissipated in the 50 Ohm termination at the 90-degree hybrid's fourth port, and a sinusoidally modulated tone has half the time averaged power of an unmodulated tone\footnote{Assuming the modulation rate is much slower than the carrier frequency.}.  Conversion gain is further reduced at certain frequencies by the presence of a chip mode around 5 GHz, impedance mismatches outside the 5-7 GHz band, and imperfect sinusoidal modulation which directs power to spurious sidebands.  Independent transmission measurements on both TIBs are shown in the supplementary material.

In addition to sideband contrast, other important specifications for a SSBM are the modulation gain, L, R, and I frequency limits, instantaneous bandwidth, linearity, and insertion losses.  %We address these in succession.
%Fig.~\ref{fig:fig3}b shows the image rejection over the same range of L and I frequencies.
Modulation gain is plotted in Fig.~\ref{fig:fig3}b as a function of $\Omega$, for tones at several different input frequencies.  Two major trends are visible: first, modulation gain decreases as $\Omega$ increases. This effect is a consequence of the bandwidth requirement of the pulse-shaping scheme used to sinusoidally modulate  transmission through the TIBs, which begins to exceed our control hardware's bandwidth as $\Omega$ increases. The second trend is a variation in the modulation gain achieved at small modulation frequencies as the probe frequency $\omega$ changes. %The second trend is a variation in the conversion gain achieved at small conversion frequencies ($\Omega \approx$ MHz), as the probe frequency $\omega$ changes.  
This is caused by a change in the phase of a TIB's transmission as its magnitude is modulated.  This effect is evident in Fig.~S2 of the supplementary material, and may be partially alleviated in a future device by removal of a chip-mode at 5 GHz. 
%Conversion gain, the power-ratio of the frequency-converted signal at the SSBM's output to the input signal, is plotted in Fig.~\ref{fig:fig3}b as a function of $\Omega$.  

%All of these are summarized in Tab.~\ref{tab:thetab}.
%\begin{widetext}
%\begin{table}[hbt]
%\centering
%\begin{tabular}{|l|l|l|l|l|l|l|}
%\hline
%R/L low       & RF/L high  & I low     & I high   & Iso. (L to R) & Image Rejection & Conversion Loss\\ \hline
%4 GHz & 8 GHz & DC & 60 MHz & 20 dB & 20 dB & 8 dB\\ \hline
%\end{tabular}
%\caption{Specifications of the single sideband modulator.}
%\label{tab:thetab}
%\end{table}
%\end{widetext}

Figs.~\ref{fig:fig3}a and~b also show the ranges of L, R, and I frequencies that can be processed by the SSBM.  The device converts frequencies from/into the 4 to 8 GHz band, shifting them by as much as 120 MHz to the red or blue.  %The upper bound on this I range is a reflection of the bandwidth limitations of the hardware used for flux control (see supplementary materials for details).  
For an FDM application in superconducting qubit readout, the quantity with which to compare %the frequency range of the I port 
this is the coupling strength of a readout cavity's strongly coupled port  $\kappa$.  Allowing 3 $\kappa$ of spectrum per readout channel, and taking a representative value of several MHz for $\kappa/2\pi$\cite{kindel:2015,kelly:2015,narla:2016,ofek:2016,hacohen:2016}, such an I range allows for 10 to 40 (or more\cite{channels}) independent readout channels per input line.  

%Ideally, the SSBM assigned to a specific qubit readout channel would convert entire wavepackets (not just pure tones) up or down in frequency.  To characterize this bandwidth, frequency conversion of amplitude modulated input tones is shown in Fig.~S4 of the supplementary information.  The instantaneous bandwidth of the SSBM exceeds 5 MHz.

%To assess device linearity, we sweep the input power during operation (data shown in Fig.~S5 of the supplementary material).  The extracted 1 dB compression point is -85 dBm, consistent with previous measurements of TIBs from the same wafer.  This power far exceeds the $-120$ dBm (or less) typically used in dispersive qubit readout.\cite{riste:2013}

Conversion of amplitude-modulated tones with the SSBM, and measurements of its linearity, are shown in supplementary Figs.~S4 and~S5.  The instantaneous bandwidth of the device exceeds 5 MHz, and its 1 dB compression point is -85 dBm, providing sufficient bandwidth and power-handling for dispersive qubit readout. 

We now discuss the insertion loss of our SSBM.  Although the dissipation of the TIBs is less than 0.5 dB\cite{chapman:2016}, our SSBM dissipates 3 dB of the input power (the entire image sideband) in the fourth port of the 90 degree hybrid at the device's output.  As noted previously, an additional 3 dB (or more) of power is dissipated/returned by our distributed network when the TIBs are operated as modulators---this is precisely the modulation gain plotted in Fig.~\ref{fig:fig3}b.  Finally, the TIBs themselves are not perfectly transmitting over the entire range of operating frequencies.  To quantify this, unnormalized transmission in the 4-8 GHz band is plotted in Fig.~S2 of the supplementary material, with independent measurements on both TIBs.  Although this transmission approaches unity at some frequencies, it is degraded elsewhere by the presence of a chip mode around 5 GHz and impedance mismatches outside the 5-7 GHz band.

A future integrated design for the SSBM could improve on these specifications: removal of the chip mode and matching the circuit impedance across the 4-8 GHz band are well-posed problems in microwave engineering.  An on-chip implementation would improve sideband contrast by suppressing path length differences in the network below a wavelength and shrink device footprint by two orders of magnitude.  An additional 3 dB of conversion gain can be recovered by using the quadrature hybrid as the power divider on the input and employing a one-to-one transformer to combine the outputs of the two TIBs---this eliminates the power dissipated in the image sideband.  A next-generation device could thus operate over a several GHz frequency range with insertion loss approaching -3 dB%, or a narrower frequency range with modulation gain approaching 0 dB if the 90 degree hybrid was replaced with a lumped-element mechanism to create the needed 90 degree phase shift between the L ports of the two TIBs
. Further considerations for a future device are provided in the supplementary material.

%Construction of a single sideband modulator from two TIBs thus provides a felicitous mechanism for frequency conversion.  
A single-sideband modulator constructed from the repeated instancing of broadband modulation elements like TIBs is thus an appealing, general purpose way to engineer frequency conversion.  The L and R bandwidths allow for conversion of tones in the 4-8 GHz band, and the device's 1 dB compression point at -85 dBm far exceeds the power typically\cite{riste:2013} used for dispersive readout.  In addition, low return, insertion, and modulation losses  are achievable in future design iterations.  Finally, %In addition, low conversion loss and robust side-band contrast are achievable in future design iterations.  Finally, %and in contrast to other cryogenic microwave frequency converters\cite{roch:2012,abdo:2013,lecocq:2016}, 
the SSBM's nonlinear elements require no GHz frequency control lines, and are actuated solely with radio frequency signals (several hundred MHz or less).  It is therefore suitable for construction of the scalable and flexible multiplexed architectures needed in future many-qubit experiments with superconducting circuits.

%We conclude by pointing out that scalable high-quality frequency conversion is a versatile tool for cyrogenic analog signal processing.  While we have highlighted applications in the FDM of superconducting qubits, other possibilities include readout of arrays of microwave kinetic inductance detectors for astronomical measurements\cite{day:2003} and in-situ tunable coupling between microwave resonators.

\vspace{0.1in}
See supplementary material for details on the experimental setup, bias-waveform pulse shaping, additional measurements, and future design considerations.

\vspace{0.1in}
\noindent{\emph {Acknowledgment}} This work is supported by the ARO under contract W911NF-14-1-0079 and the National Science Foundation under Grant Number
1125844.  The authors thank Bradley A. Moores and Andrew P. Higginbotham for helpful discussions.

\pagebreak
\widetext
\begin{center}
\textbf{\large Supplementary Material for\\ ``Single-sideband modulator for frequency domain multiplexing of superconducting qubit readout''}
\end{center}
%%%%%%%%%% Merge with supplemental materials %%%%%%%%%%
%%%%%%%%%% Prefix a "S" to all equations, figures, tables and reset the counter %%%%%%%%%%
\setcounter{equation}{0}
\setcounter{figure}{0}
\setcounter{table}{0}
\setcounter{page}{1}
\makeatletter
\renewcommand{\theequation}{S\arabic{equation}}
\renewcommand{\thefigure}{S\arabic{figure}}
\renewcommand{\bibnumfmt}[1]{[S#1]}
\renewcommand{\citenumfont}[1]{S#1}

\section{Experimental setup for measurement of a single-sideband modulator}
\begin{figure}[htb]
\begin{center}
\includegraphics[width=0.8\linewidth]{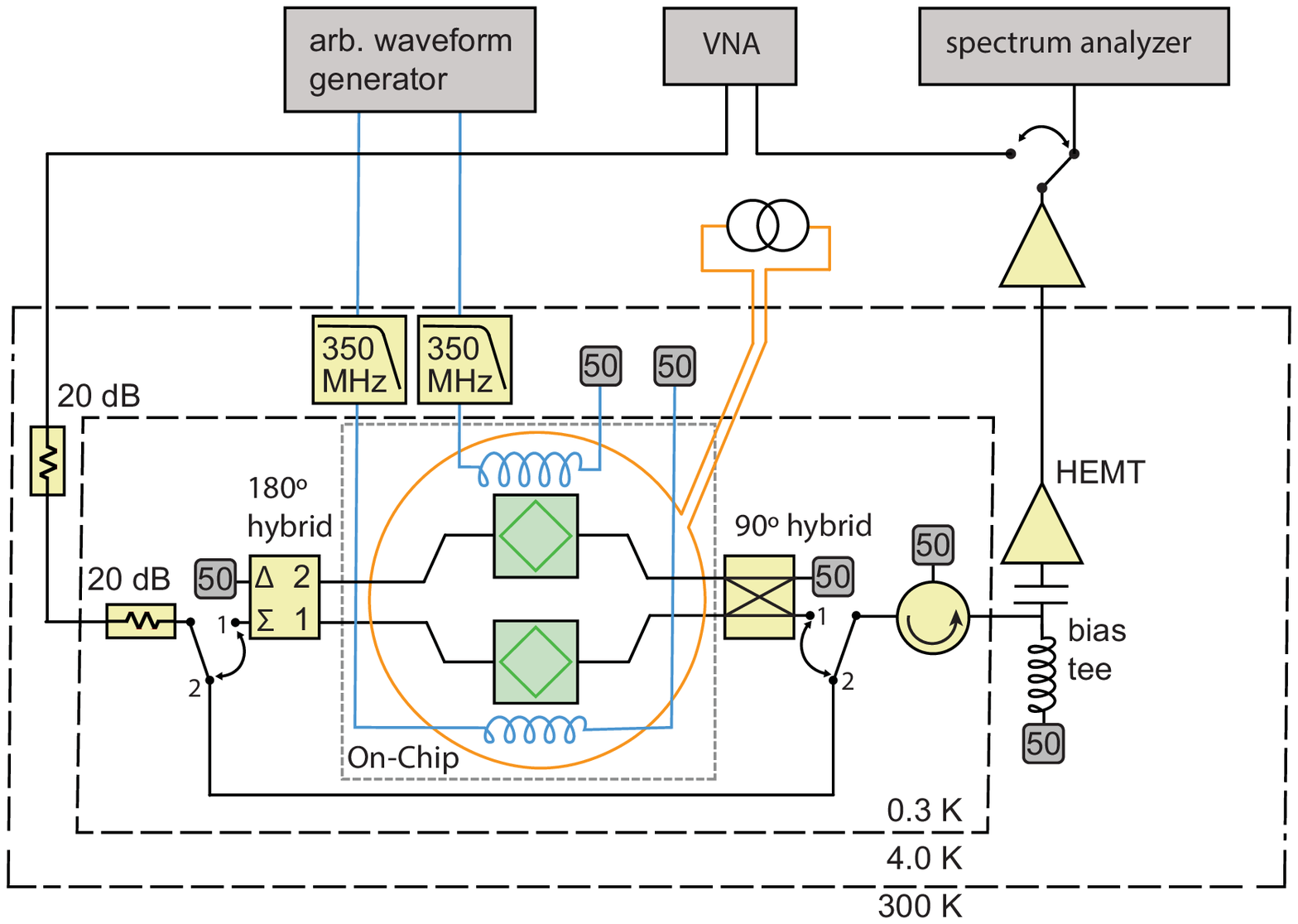}
\caption {\small Wiring diagram for measurements of the single-sideband modulator.  TIBs are represented by green diamonds.  The orange circle represents an off-chip magnetic coil, and blue inductor symbols represent on-chip gradiometric flux control lines.}
\label{fig:supp1}
\end{center}
\end{figure}

A schematic of the experimental setup is shown in Fig.~\ref{fig:supp1}.  A microwave generator creates a tone that is routed into a $^3$He cryostat and attenuated.  The tone impinges on the sum port of a 180 degree hybrid coupler (Krytar 4040124), which divides the power into two coherent components.  The signals then travel through the two TIBs, and are summed together with a 90 degree hybrid (Fairview Microwave SH7219).  Finally, the tone propagates through the microwave receiver and is readout in a spectrum analyzer or vector network analyzer.

To control the transmission of the TIBs, three control currents are used.  First, an off-chip magnetic coil (orange circle) creates a uniform magnetic flux $\phi_{\Sigma}$ across the entire chip ($|H| \approx 300$ mOe).  The second and third control currents run in a pair of on-chip bias lines (blue inductors) that create gradiometric fluxes for the two TIBs.  To operate a TIB as a double-balanced modulator, we fix the current in the off-chip magnetic coil and dynamically modulate the current $I$ in its on-chip bias line.  This changes the imbalance of the TIB, and thus its transmission $T$.  %We denote the transmission through a TIB as $T(t)$.  The time dependence in this quantity comes from the implicit time dependence in its bias current.

\section{Pulse shaping control current waveforms to optimize modulator performance}
Judicious choice of this control current $I$ can substantially improve modulator performance.  To calibrate this, we first use a network analyzer to measure the transmission through each TIB as a function of an applied DC current in its on-chip bias line.  The other TIB is unbiased during this measurement (its inductor bridge is balanced and incident signals will be reflected) allowing for independent calibration of both TIBs.

\begin{figure}[htb]
\begin{center}
\includegraphics[width=1.0\linewidth]{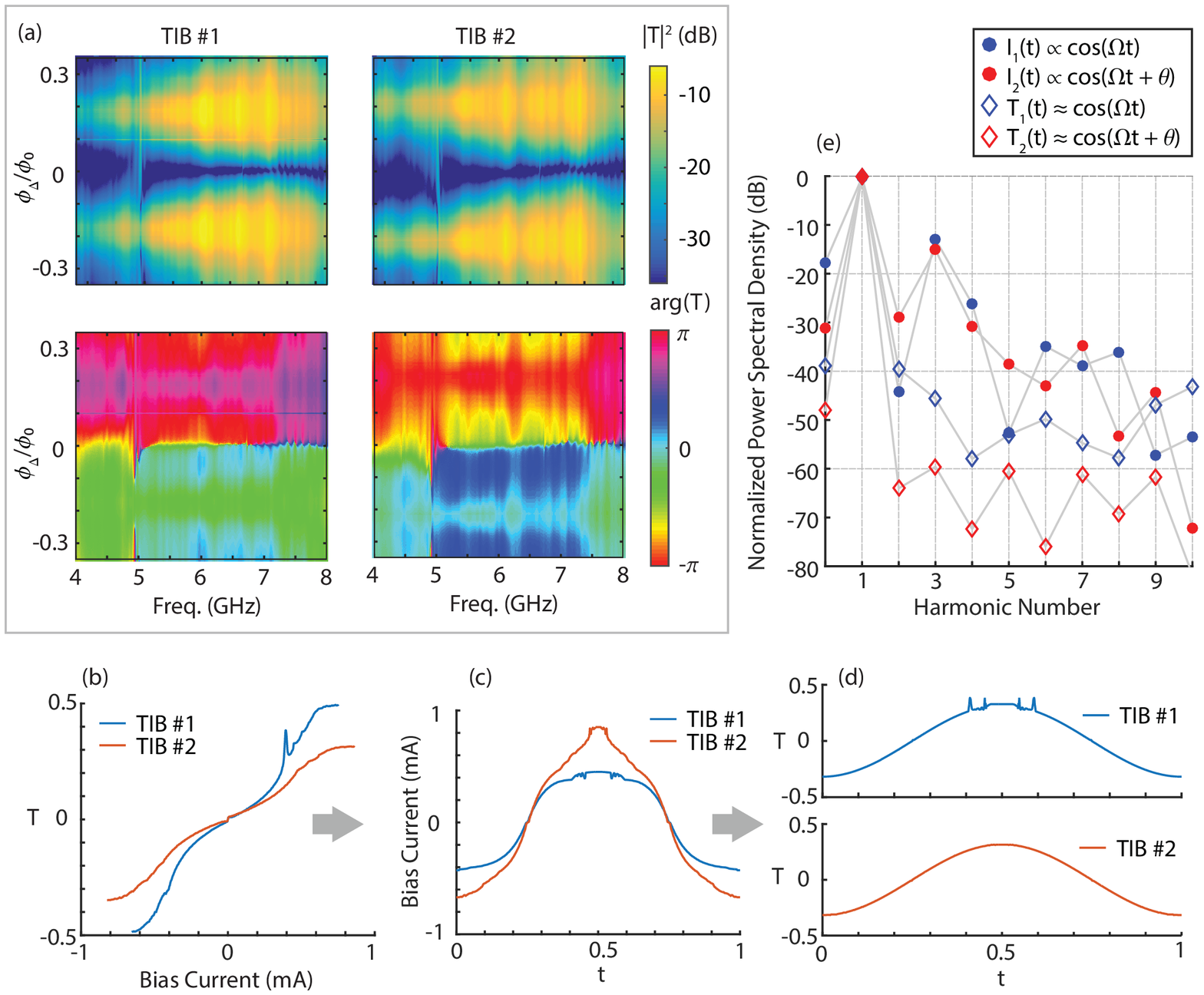}
\caption {\small (a) Amplitude and phase of transmission measurements on each TIB, while the other TIB is set to reflect incident signals (zero bias current). The maximum of the transmission color scale is -6 dB. The phase of transmission shifts by approximately 180 degrees as the bias current changes polarity. (b) Linecuts of transmission vs bias current $T(I)$, with probe frequency fixed at 6 GHz. When bias current is swept sinusoidally, transmission changes in a non-sinusoidal manner. (c) Applying a bias current $I(t) = T^{-1}(\cos(\Omega t))$ produces approximately sinusoidal modulation of transmission, as seen in (d). For TIB \#1, the map $T(I)$ is not invertible, which causes some distortion (higher harmonics of $\Omega$) in the transmitted spectrum. (e) Power spectral density of transmission when current is modulated sinusoidally (filled circles), and when the current is chosen to produce sinusoidal transmission (unfilled diamonds). Spectra are normalized to the total power in each sideband. Pulse shaping reduces the weight of unwanted higher harmonics by approximately 20 dB.  Data for TIB\#1 (TIB\#2) are shown in blue (orange).}
\label{fig:supp2}
\end{center}
\end{figure}

Fig.~\ref{fig:supp2}a shows the squared amplitude (top row) and phase (bottom row) of signals transmitted through each TIB, as a function of the probe frequency and a gradiometric flux $\phi_\Delta$ that is proportional to the control current.  In the squared amplitude plots, one observes low transmission when $\phi_{\Delta}/\phi_0$ is small, and maximal transmitted power when $\phi_{\Delta}/\phi_0 = 1/5$.  Here $\phi_0 = \hbar/2e$ is the reduced flux quantum, with $\hbar$ Planck's constant over 2$\pi$ and $e$ the charge of an electron.  The uniform flux is fixed at $\phi_{\Sigma} = 3 \phi_0/10$.  Because half the signal is dissipated in the 50 Ohm termination on the 90 degree hybrid, and another half is reflected by the un-probed arm of the network, the maximum possible transmission in this configuration is -6 dB.  From the plots of phase in the bottom row, one can see that inverting the polarity of the bias current causes a $\pi$ phase shift in the transmission. 

Linecuts of the real part of the transmission at 6 GHz are shown in Fig.~\ref{fig:supp2}b.  These are maps $T(I)$ between the applied control current and the transmission through the bridges.  If these maps were linear, a sinusoidal control current would induce a sinusoidal modulation of the transmission through the TIB.  Transmission, however, is a nonlinear function of the control current.\cite{schapman:2016}  To ensure sinusoidal modulation by the TIBs at the desired angular frequency $\Omega$, we use an arbitrary waveform generator to apply the control current $I(t) =  T^{-1}(\cos(\Omega t))$ (Fig.~\ref{fig:supp2}c).  Here $T^{-1}$ denotes the inverse of the map $T(I)$.  The time-dependence of the transmission, $T(t)$, is then ensured by function composition to be the desired value of $\cos(\Omega t)$ (Fig.~\ref{fig:supp2}d).  When $T(I)$ is not a one-to-one mapping (Fig.~\ref{fig:supp2}b, blue curve) this procedure only produces an approximation of sinusoidal modulation, as seen in Fig.~\ref{fig:supp2}d, top panel. (As the optimal bias current waveform $I(t)$ must be sampled discretely before loading onto the arbitrary waveform generator and is bandwidth limited by that arbitrary waveform generator, in practice $T(t)$ is a smoother function than that shown in Fig.~S2d.) 
The overall effect of this procedure can be seen in Fig.~\ref{fig:supp2}e, which plots the power spectral density of $T(t)$ for both TIBs, when the bias currents are sinusoidal (circles) and when the bias currents are chosen to produce sinusoidal transmission (diamonds).  The latter method reduces the spectral weight of higher harmonics by about 20 dB. %This can be seen in Fig.~\ref{fig:supp2}e, which plots the power spectral density of $T(t)$ for both TIBs, when the bias currents are sinusoidal (circles) and when the bias currents are chosen to produce sinusoidal transmission (diamonds).  The latter method reduces the spectral weight of higher harmonics by about 20 dB.

\section{IF frequency limitations}
\begin{figure}[htb]
\begin{center}
\includegraphics[width=1.0\linewidth]{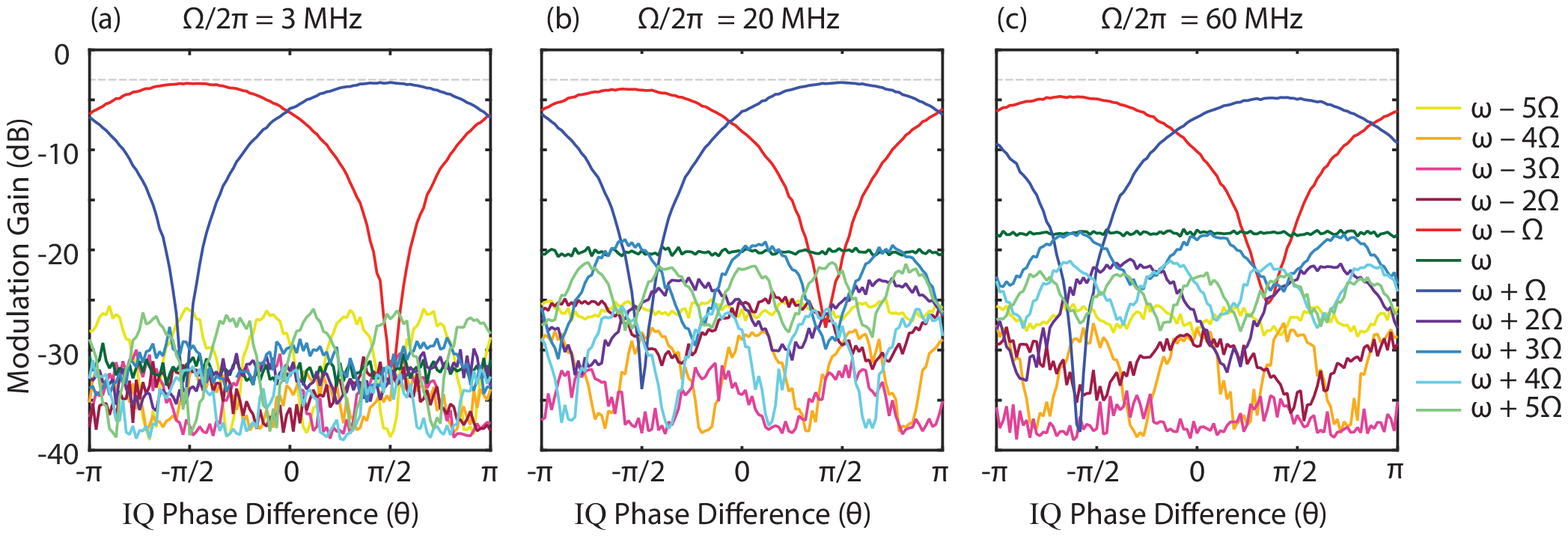}
\caption {\small Sideband power as a function of IQ phase difference for modulation rates of $\Omega/2\pi = $ 3 MHz (a), 20 MHz (b), and 60 MHz (c), with $\omega/2\pi = 4$ GHz. A gray dashed line at -3 dB shows the expected upper-bound on transmitted power.}
\label{fig:supp3}
\end{center}
\end{figure}

Harmonics in the spectrum of $T(t)$ are imprinted directly onto the spectrum of signals transmitted through the SSBM.  Fig.~\ref{fig:supp3} shows the nearest 10 harmonics around as the IQ phase difference is swept.  The LO port of the device is driven by a microwave tone at $\omega = 2 \pi \times 4$ GHz and the I and Q ports are driven by angular frequencies of $\Omega = 2 \pi \times 3$ MHz (Fig.~\ref{fig:supp3}a),  $\Omega = 2 \pi \times 20$ MHz (Fig.~\ref{fig:supp3}b), and  $\Omega = 2 \pi \times 60$ MHz (Fig.~\ref{fig:supp3}c).

As $\Omega$ is increased and signals are converted farther from the input frequency, the output at both the input frequency and at spurious harmonics increase in magnitude.  This is a consequence of the finite bandwidth of our arbitrary waveform generator (Tektronix AWG5014c), which samples at 1.2 Gs/s and has a bandwidth of 180 MHz.  (For input powers much less than the 1 dB compression point, the size of spurious harmonics does not depend on the power of the input signal.)  As the IF modulation rate $\Omega$ grows, the number of points available to represent the control current pulse $I(t)$ decreases, and the bandwidth needed to capture fine corrections to $I(t)$ increases.  This diminishes the ability of the pulse-shaping routine to ensure approximately sinusoidal transmission.  For instance, as $\Omega/(2 \pi)$ changes from 3 to 60 MHz, the image rejection goes from 30 dB to 20 dB, and the sideband contrast goes from 25 dB to 13 dB.

%Increasing $\Omega$ also reduces conversion gain.  Fig.~\ref{fig:convgain} shows this effect, for different LO frequencies.
%\begin{figure}[htb]
%\begin{center}
%\includegraphics[width=1.0\linewidth]{SuppFigures/Supp5.eps}
%\caption {\small Conversion gain as a function of modulation rate, $\Omega$, plotted for different probe frequencies. Solid lines refer to the 1$n\textsuperscript{st}$ lower sideband and dashed lines refer to the 1$n\textsuperscript{st}$ upper sideband.}
%\label{fig:convgain}
%\end{center}
%\end{figure}
%A future version of the SSBM could increase conversion gain to -6 dB for all the plotted LO and IF frequencies.  We discuss design considerations for such a device shortly.

\section{Frequency conversion of amplitude-modulated microwaves}
\label{sec:IBW}
The previous sections shows frequency conversion of pure microwave tones with an SSBM.  In the envisaged application, however, the SSBM must convert wavepackets with more complex spectral composition.  A realistic example of such a signal is a microwave tone that is phase modulated in time, at a rate set by the bandwidth of the qubit readout cavity.  Such a signal simulates the dataflow from a qubit-cavity system that is being continuously monitored, say, for the detection of quantum jumps: as the qubit state stochastically jumps between ground and excited, the phase of the transmitted microwave tone concomitantly acquires a phase shift of 0 or $\pi$ radians. The phase modulation of the transmitted signal creates a wavepacket with a richer frequency content than the orginal tone, and extraction of the information in the phase modulation requires single-sideband modulation of the entire wavepacket.

\begin{figure}[htb]
\begin{center}
\includegraphics[width=1.0\linewidth]{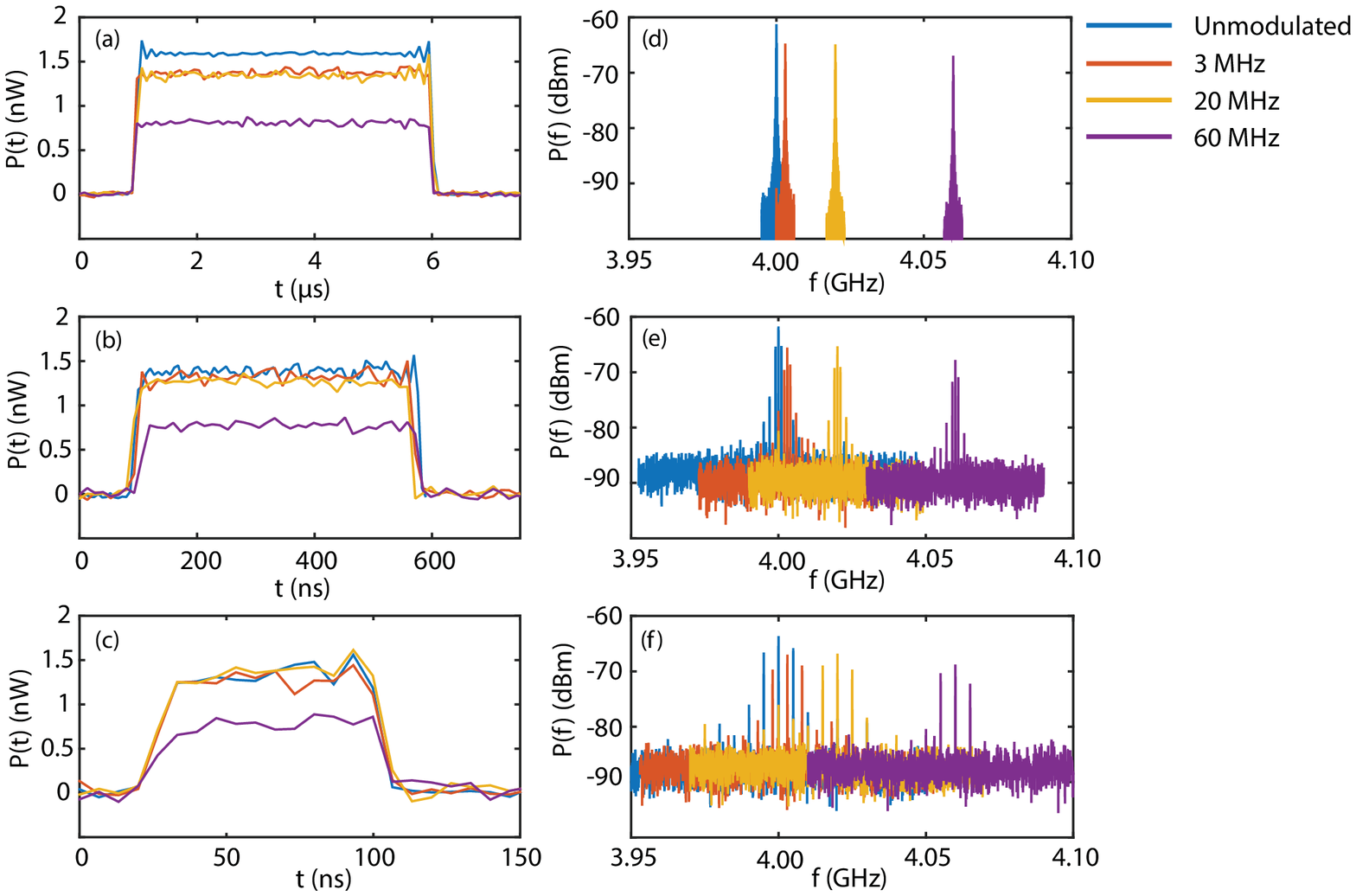}
\caption {\small Amplitude modulation of an $\omega = 2 \pi \times 4$ GHz tone after frequency conversion by the SSBM.  Transmitted power versus time (a-c) and power spectral density (d-f) are shown for modulation rates of 100 KHz (top row), 1 MHz (middle row), and 5 MHz (bottom row).  At each of the three modulation rates, we use the SSBM to convert the frequency of an input tone by 0, 3, 20, and 60 MHz.  These are depicted by the four traces in each panel.  For visual clarity, the amplitude of the unconverted signals (blue traces) in (a-c) are scaled by a factor of 1/2.}
\label{fig:supp4}
\end{center}
\end{figure}

To simulate this process, we use the SSBM to frequency-convert a microwave tone that is amplitude modulated with a square wave at frequency $\nu$.  (For this demonstration, we use amplitude modulation in place of phase modulation to allow detection with a spectrum analyzer.)  We measure the modulation of the signal's amplitude in time (Fig.~\ref{fig:supp4}a-c), and the spectrum of the converted signal (Fig.~\ref{fig:supp4}d-f).  Amplitude modulations of $\nu = $ 100 KHz, 1 MHz, and 5 MHz are shown in the top, middle, and bottom rows of the figure.  For each modulation, the four traces show frequency conversion by 0, 3, 20, or 60 MHz, as evidenced by the shifted spectra in the figure's right column.

From Fig.~\ref{fig:supp4}, one can see that the instantaneous bandwidth of the SSBM exceeds 5 MHz.  At higher frequencies, for example 5 MHz, distortion of the square-wave begins to become apparent.  We believe the instantaneous bandwidth of the device is currently limited by the relaxation of eddy currents in the chip ground plane.  This relaxation time can be substantially reduced in future devices.  

\section{Linearity}
\begin{figure}[htb]
\begin{center}
\includegraphics[width=0.5\linewidth]{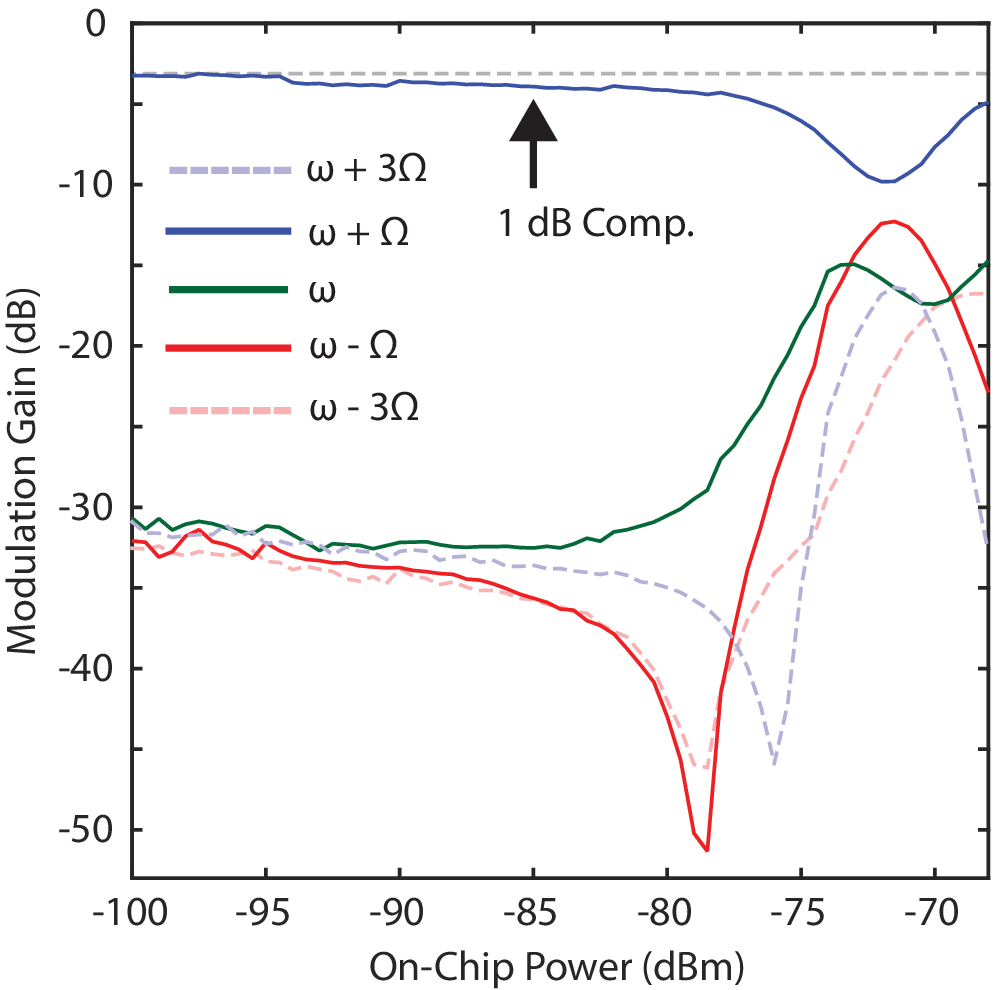}
\caption {Selected sideband power as a function of on-chip power, with $\omega/2\pi = 4$ GHz, $\Omega/2\pi$ = 3 MHz, and $\theta = 90$ degrees. 1 dB compression occurs at -85 dBm. Dashed line at -3 dB is a guide to the eye.%The dashed gray lines has a slope of three, demonstrating the linearity of our device below the 1 dB compression point and the nonlinearity of the device above. %The IP3 point is about -64 dBm.
}
\label{fig:linearity}
\end{center}
\end{figure}

To assess the power-handling of the SSBM we sweep the on-chip power during operation at $\Omega/2\pi = 3$ MHz, $\omega/2\pi = 4$ GHz, and $\theta = 90$ degrees (up-conversion), and measure the power transmitted at various harmonics of the modulation rate.  Fig.~\ref{fig:linearity} shows the results.  We measure the power in the nearest 18 harmonics $(n\leq 9)$ to the input frequency. For visual clarity, only the zeroth and the four lowest odd harmonics are plotted, though the other harmonics exhibit similar behavior.

When the incident power is below -95 dBm, the output power at the input frequency and in all the sidebands scales linearly.  As the input power increases, 1 dB compression in the first blue sideband occurs at -85 dBm.  This value is consistent with previous results, which found the 1 dB compression point of a single TIB to be -88 dBm~\cite{schapman:2016}.  (As the TIBs in our SSBM need only process one-half the power incident on the device's input, the SSBM's compression point is 3 dB below that of a single TIB.) In a dispersive readout scheme the power of a readout tone depends on the dressed cavity bandwidth, but most tones fall between -130 and -120 dBm~\cite{sriste:2013}. The SSBM's 1 dB compression point at -85 dBm thus provides ample power handling for this application. 

\section{Considerations for a future, on-chip, integrated design}
As the SSBM is created with two instances of a TIB, design improvements for the single-sideband modulator may be partitioned into two categories: changes of the TIB, and changes of the composite device (the two TIBs, hybrid couplers, and internal microwave connections).  We address these in turn.

As mentioned in the main text, immediate improvements on the TIB may be realized by eliminating the chip-mode around 5 GHz (apparent in Fig.~\ref{fig:supp2}a), and by improving the impedance match.  The chip-mode is an LC resonance between the inductance of the balun coils and a parasitic capacitance to ground, which can be tuned out of the operation band by manipulating the width of the balun microstrips (their length is constrained to be a quarter wavelength at the operation frequency).  

In the current generation of devices, impedance matching a TIB to 50 Ohms is inhibited by the finite tunability of the inductors in the bridge and the impedance of those inductors (about 20 Ohms at 6 GHz) with respect to 50 Ohms.  We overcome this problem by tuning out the residual inductance with capacitors in series with the four nodes of the bridge.  Although this is an expedient solution, the resulting match is effective only within several hundred MHz of the frequency for which the capacitors have the appropriate impedance.  A more sophisticated approach, such as tapered lines or matching transformers, could easily expand this bandwidth to several GHz, as the Bode-Fano criterion~\cite{bode:1945,fano:1950} is relatively lenient for low $Q$ circuits. More quantitatively, an imbalanced TIB may be modeled as an inductor in series with a 50 Ohm load.  The Thevenin equivalent inductance of a TIB with inductors parametrized as $l_1$ and $l_2$ (as in Fig.~2b of the main text) is $2 l_1 l_2 / (l_1 + l_2)$, which for the TIBs in this device amounts to 800 pH when the bridges are maximally imbalanced. The Bode-Fano criterion thus allows for -20 dB reflections over an 8 GHz bandwidth.  Improved impedance matching may therefore drastically reduce return loss over the 4 to 8 GHz band.

Better impedance matching can also be achieved directly by improving the tunability (dynamic range) of the SQUID critical currents.  While reducing the number of SQUIDs in each array will accomplish this by reducing geometric inductance and enhancing participation ratios, power handling will suffer.  Another way to improve tunability is to more uniformly distribute the gradiometric flux across the length of the array, which reduces variation in the critical currents of the SQUIDs that make up the array.  In the current layout, SQUIDs near the inside corners of the bias line couple more strongly to it, and are threaded by greater amounts of flux.  Laying out the figure-eight of the bridge so that the bias line can bisect it without making turns will alleviate this problem, increase the degree to which the bridge can be imbalanced, and reduce impedance mismatch.

Finally, the instantaneous bandwidth of the TIB may also be improved by elimination of  long timescale eddy currents in the chip ground plane.  This can be done by altering the resistance (modifying the $L/R$ time) of a normal metal layer placed in the chip ground plane to break supercurrent loops.

In addition to improving the constituent components, a next-generation SSBM can also benefit from complete on-chip integration.  First, on-chip superconducting hybrids can be extremely low-loss.  Second, path length differences in the two arms of the network can be easily constrained to micron length scales, making them smaller than the operation wavelengths by more than four orders of magnitude. (To contrast, in our current device this path length difference is 3 mm).  Suppressing path length difference far below the operation wavelength allows for simultaneous minimization of the power in the image sideband and maximization of the power in the desired sideband.

A third considerable benefit of moving to a completely monolithic circuit is the recovery of an additional 3 dB of modulation gain.  In the current implementation, an off-chip power splitter divides the input power and delivers it to the two on-chip TIBs.  During operation, this distributed network suffers from return loss, as each TIB is mismatched from 50 Ohms during portions of the modulation period (a portion of this reflected power is dissipated in the power divider at the input, and another portion exits the device's input as return loss).  It also results in an additional 3 dB of insertion loss; when signals in the two arms of the interferometer are summed in the 90 degree hybrid, the image (undesired) sideband is dissipated in the terminated port of the hybrid.  

To remedy this, we propose using a 90-degree hybrid coupler as the power-divider at the device's input, and combining the signals in the interferometer with a one-to-one transformer.  Although this layout does not recover the 3 dB of return and insertion loss at the device's input, it eliminates the 3 dB of insertion loss from the image sideband at the output.  An SSBM designed in this way could operate with a modulation gain of -3 dB.  %As discussed in the main text, the remaining 3 dB of conversion gain could be recovered by moving to a fully lumped-element design including the removal of the 90 degree hybrid.  Such a move, however, would require an alternative method of producing the required 90 degree phase shift, which may restrict the operation bandwidth. 


\begin{thebibliography}{36}
\bibitem [{cha()}]{channels}%
  \BibitemOpen
  \href@noop {} {\ }\bibinfo {note} {More readout channels are possible if the
  cavities in each channel have distinct dressed frequencies or lower cavity
  coupling rates.}\BibitemShut {Stop}%
\end{thebibliography}
\end{document}